\documentclass[referee]{raa}   
\usepackage{graphicx,times}
\usepackage{natbib}
\usepackage{amssymb,amsmath}
\usepackage{lineno}
\bibpunct{(}{)}{;}{a}{}{,}
\usepackage[pagebackref=true]{hyperref}

\newcommand\M{\rm M}
\newcommand\dd{\rm d}
\newcommand\p{\rm p}
\newcommand\fid{\rm fid}

\begin{document}
\title{Dynamical dark energy in light of cosmic distance measurements I: a demonstration using simulated datasets}

 \volnopage{ {\bf 20XX} Vol.\ {\bf X} No. {\bf XX}, 000--000}
   \setcounter{page}{1}

   \author{Gan Gu
   \inst{1,2}, Xiaoma Wang\inst{1,2}, Xiaoyong Mu\inst{1,2}, Shuo Yuan
      \inst{1},  Gong-Bo Zhao\inst{1,2,3}
   }

   \institute{National Astronomical Observatories, Chinese Academy of Sciences, Beijing, 100101, P.R.China, 
China; {\it gbzhao@nao.cas.cn}\\
    \and
             University of Chinese Academy of Sciences, Beijing, 100049, P.R.China\\
	\and
Institute for Frontiers in Astronomy and Astrophysics, Beijing Normal University, Beijing, 102206, P.R.China\\
\vs \no
   {\small Received 2024 Month Day; accepted 2024 Month Day}
}

\abstract{We develop methods to extract key dark energy information from cosmic distance measurements including the BAO scales and supernovae luminosity distances. Demonstrated using simulated datasets of the complete DESI, LSST and Roman surveys designed for BAO and SNe distance measurements, we show that using our method, the dynamical behaviour of the energy, pressure, equation of state (with its time derivative) of dark energy and the cosmic deceleration function can all be accurately recovered from high-quality data, which allows for robust diagnostic tests for dark energy models. 
}

   \authorrunning{G. Gu et al. }   
   \titlerunning{Probing dark energy using simulated datasets}  
   \maketitle

\section{Introduction}         
\label{sec:intro}

The physical origin of the accelerating expansion of the Universe, which was discovered using observations of type Ia supernovae in $1998$ \citep{1998AJ....116.1009R, 1999ApJ...517..565P}, remains unknown. Possible mechanisms for the cosmic acceleration include dark energy \citep{DE}, which is a dominating component of the cosmic energy budget today with a negative pressure, and modified gravity \citep{MG}, a framework in which Einstein's general relativity (GR) gets modified. In both scenarios, an effective equation of state of dark energy, $w$, defined as a ratio of the pressure $P$ over energy density $\rho$ of the effective dark fluid, is a critical quantity for investigating models that can explain the cosmic acceleration. For example, $w=-1$ may mean that dark energy is essentially the vacuum energy, while an evolving $w$ with cosmic time may suggest the dynamical nature of dark energy, or a breakdown of GR on cosmic scales. Therefore a direct reconstruction of $w$ as a function of the scale factor $a$ (or redshift $z$) from observations is of general interest \citep{DErecon1,DErecon3,DErecon4,DErecon2,DErecon5,DErecon6,DErecon7}. 

However, a reconstruction of $w(a)$ is not straightforward. Parametric reconstructions are easier to perform given the small number of free parameters to be determined, but the resultant reconstruction may be biased as it can only take the functional form assumed in the first place, which may not be appropriate. Non-parametric reconstructions are more general, but given the large number of degrees of freedom in the process, various kinds of datasets, such as the cosmic microwave background (CMB) \citep{WMAP,Planck18}, type Ia supernovae (SNe Ia) \citep{1998AJ....116.1009R, 1999ApJ...517..565P}, baryonic acoustic oscillations (BAO) \citep{BAO98,BAO05a,BAO05b,eBOSS:2020yzd} and redshift space distortions (RSD) \citep{Kaiser,Peacock:2001gs,eBOSS:2020yzd}, are combined for the purpose of degeneracy breaking. If one or more kinds of datasets are contaminated by unknown systematics, which is not impossible, the final reconstructed $w(a)$ inherits the systematics. In this sense, it is better to learn $w(a)$ from each individual type of datasets, and cross check the consistency, especially when datasets are in tension. But unfortunately this is difficult for non-parametric reconstruction methods.

Actually, one can learn important features of $w(a)$ without a direction reconstruction. In this work, we derive useful diagnostic quantities for $w(a)$ from cosmic distance measurements including the BAO scales and SN Ia luminosity distances, and validate our tests using simulated datasets including the galaxy survey of Dark Energy Spectroscopic Instrument (DESI) \citep{DESI:2016fyo} and SNe Ia surveys of Rubin Observatory’s Legacy Survey of Space and Time (LSST) \citep{LSST} and the Roman Space Telescope \citep{Roman}. 

This paper is structured as follows. We develop
the methodology in Sec. 2, describe the simulated datasets in Sec. 3, and present the main result in Sec. 4, before conclude in Sec. 5. Some technical details are included in the Appendices.

\section{Features of dark energy hidden in distance measurements} \label{sec:method}

In this section, we show the information content of cosmic distance measurements that is relevant for dark energy studies, and propose methods to extract this piece of crucial information.

\subsection{The shape function of dark energy}

In a spatially flat Universe, the Hubble expansion rate $H(a)$ is related to the fractional dark energy density $X(a)$ through \begin{equation}\label{eq:H2a3}
f(a) \equiv A H^2 a^3=B \left[X(a) a^3\right]+C,
\end{equation} where $A,~B$ and $C$ are constants. For example, for BAO observables, $A=r_{\dd}^2$, $B=r_{\dd}^2 H_{0}^{2}(1-\Omega_{\M}), C=r_{\dd}^2 H_{0}^{2}\Omega_{\M}$ with $H_0, r_{\dd}$ and $\Omega_{\M}$ being the Hubble constant, the sound horizon at decoupling, and the present-day fractional matter density, respectively. The quantity $X(a)$ is defined as,
\begin{equation}\label{eq:Xrho}
   X(a) \equiv \frac{\rho_{\mathrm{DE}}(a)}{\rho_{\mathrm{DE}}(a = 1)} = {\rm exp} \left [-3\int_{1}^{a}\frac{1+w(y)}{y}{\dd} y \right ],
\end{equation} where $\rho_{\rm DE}$ and $w$ are the mean energy density, and the equation of state of dark energy, respectively.

From Eq. (\ref{eq:H2a3}), it is clear that functions $H^2 a^3$ and $Xa^3$ have the same shape, meaning that they are identical after a proper shift and normalisation. For example, \begin{equation} \label{eq:fNorm}
 S[A H^2 a^3]= S[X a^3]; \ \ S[f(a)] \equiv \frac{f(a)-f(a_{\star})}{f'(a_{\star})},
\end{equation} where $S[f(a)]$ defines a shape function of $f(a)$, and $a_{\star}$ is a reference scale factor. Throughout the paper, the superscript $'$ denotes a derivative with respect to the scale factor $a$. Although the choice of $a_{\star}$ is arbitrary, it makes sense to choose one so that $f'(a_{\star})$ can be well measured, thus we can get a decent estimation for $S$ and other quantities that depend on $f'(a_{\star})$. For the simulated datasets used in this work, we find that $a_{\star}=2/3$ is a reasonable choice to yield a tight constraint of $f'(a_{\star})$ using either the BAO or SNe datasets, so we set $a_{\star}=2/3$ for all results in this paper. We can obtain the shape information of $Xa^3$ through $S[A H^2a^3]$, which is a direct observable, and can be used as a diagnostic for dark energy models. For example, \begin{equation}
\Lambda{\rm CDM} \Longrightarrow S[A H^2a^3]=\frac{a^3-a_{\star}^3}{3a_{\star}^2 }.
\end{equation} 

\subsection{The pressure of dark energy}
\label{sec:P-function}

Since the pressure $P$ of dark energy is proportional to $wX$, it follows that

\begin{equation}
\label{eq:Rdef}
    R(a,a_{\star})\equiv\frac{P(a)}{P(a_{\star})} 
    = \frac{w(a)X(a)}{w(a_{\star})X(a_{\star})}
    =\frac{a_{\star}^2}{a^{2}}\frac{f^{\prime}(a)}{f^{\prime}({a_{\star}})}
\end{equation}
where $a_{\star}$ denotes a reference point for a normalization. From the definition, it is clear that,
\begin{eqnarray}
\label{eq:Rmodel}
w=-1 & \Longrightarrow & R=1,\nonumber \\ 
w={\rm constant} & \Longrightarrow & R= \left(\frac{a}{a_{\star}}\right)^{-3(1+w)}\Longrightarrow \frac{{\rm log}~R}{{\rm log}(a/a_{\star})}={\rm constant}
\end{eqnarray}

\subsection{The characterisation function of dark energy}
\label{sec:g-function}

To obtain further information of $w(a)$, we take second derivative of $f(a)$ with respect to $a$ and compare it to $f^\prime(a)$\footnote{An explicit derivation is included in Sec. \ref{sec:g}.}. Specifically,
\begin{equation}\label{eq:g}
    g(a) \equiv -\frac{1}{3}\left[a\frac{f^{\prime\prime}(a)}{f^{\prime}(a)} + 1\right]= w-\frac{a}{3} \frac{w^{\prime}}{w} . 
\end{equation}
This function is a direct observable from distance measurements \citep{Statefinder1,Statefinder2}, and it contains only $w$ and $w^\prime$. This means that it is free from degeneracies with any other cosmological parameters such as $\Omega_{\M}, H_0$, {\it etc}. From the definition of $g(a)$, we see that,
\begin{eqnarray}
w=-1 & \Longrightarrow & g=-1, \nonumber\\
w={\rm constant} & \Longrightarrow & g=w={\rm constant,} \nonumber\\ 
{\rm At~high~redshifts}  & \Longrightarrow & g\simeq w, \nonumber\\ 
w~{\rm slowly~varies~with~time} & \Longrightarrow & g\simeq w, \nonumber\\
w=w_0+w_a(1-a) & \Longrightarrow & g~{\rm is~close~to~a~linear~function~of}~a.
\end{eqnarray} These features can be used as diagnostics for dark energy models. For example, $g\ne-1$ rules out the $\Lambda$CDM model, and a varying $g$ with time rules out the $w$CDM model (the model in which $w$ is a constant). Furthermore, given a measurement of $g(a)$, we can obtain a relation between $w$ and $w^\prime$ at any redshifts. This can in principle be used to differentiate dark energy models in the $w-w^\prime$ phase-space \citep{w-wp1,w-wp2,w-wp3}, which is presented in a companion paper \citep{XMW24}.

\subsection{The deceleration function}\label{sec:q-function}

The deceleration function $q(a)$ is another useful quantity that can be derived from $f(a)$, although it is not solely depends on dark energy parameters. Specifically,
\begin{equation}\label{eq:q-rec}
    q(a) \equiv -a\frac{H^{\prime}}{H}-1 
    =\frac{1}{2}\left[1 - \frac{af^{\prime}(a)}{f(a)}\right].
\end{equation}

\subsection{The parametrisation of the cosmic distances}
\label{sec:f_rec}

All the above-mentioned useful functions can be derived from $f(a)$, thus it is important to derive $f(a)$ from distance measurements in an efficient and accurate way. Following \cite{Zhu_2015}, we parametrise the cosmological distances in form of,
\begin{eqnarray}
\label{Eq:para-DA}
    \frac{D_{\rm A}(a)}{D_{\rm A, fid}(a)}&=&\alpha_{0}\left(1+\alpha_{1}x+
    \frac{1}{2}\alpha_{2}x^{2}+
    \frac{1}{6}\alpha_{3}x^{3}+
    \frac{1}{24}\alpha_{4}x^{4}+    
    \frac{1}{120}\alpha_{5}x^{5}+
    \frac{1}{720}\alpha_{6}x^{6}\right),\nonumber\\
x&\equiv&\frac{\chi_{\rm fid}(a)}{\chi_{\rm fid}(a_{\rm p})} - 1, 
\end{eqnarray}
where $a_{\p}$ is the pivot point for the expansion, and we choose to set $a_{\p}=2/3$ in this work. As demonstrated in Sec. \ref{sec:zp}, the choice of $a_{\p}$ has almost no impact on the final reconstruction result. Since Eq. (\ref{Eq:para-DA}) is essentially a Taylor expansion, we need a criterion to determine the maximal order of expansion to be included in the series. Keeping more terms in the expansion makes this parametric reconstruction more accurate, but this also inflates the uncertainties due to the degeneracies among parameters. Therefore a balance between the reconstruction bias and the statistical uncertainties is required when determining the highest order of the expansion. Note that this depends on the dataset being used - better measured data can help to constrain more parameters. Given the sensitivity of DESI, LSST and Roman, we find that keeping $\alpha_0$ to $\alpha_6$ terms in the expansion is a sensible choice when all datasets are combined\footnote{For the case of BAO alone, we keep $\alpha_0$ to $\alpha_4$ terms to avoid overfitting.} for the four fiducial dark energy models considered in this work (see a demonstration in Sec. \ref{sec:order}), which cover a wide range of phenomenological dark energy models.

The subscript `fid' stands for the fiducial cosmology, which is chosen to be a $\Lambda$CDM model favored by the latest Planck observations \citep{2020A&A...641A...6P}. Using the relation between $H$ and $D_{\rm A}$ in a flat Universe, we find that, \begin{equation}\label{Eq:para-H}
    \frac{H_{\rm fid}(a)} {H(a)} =\beta_{0}\left(1 +\beta_{1}x+
    \beta_{2}x^{2}+
    \beta_{3}x^{3}+
    \beta_{4}x^{4}+
    \beta_{5}x^{5}+
    \beta_{6}x^{6}\right),
\end{equation} where parameters $\beta_{i}$ can be derived from $\alpha_{i}$, and an explicit derivation is included in Appendix \ref{sec:expansion}. Equations (\ref{Eq:para-DA}) and (\ref{Eq:para-H}) allow for a parametric reconstruction of $f(a)$ from distance measurements\footnote{Note that when fitting $\alpha$'s in Eqs. (\ref{Eq:para-DA}) and (\ref{Eq:para-H}) to distance measurements, the derived $X$ may not be positive-definite. Therefore when deriving quantities related to $f'$ including the pressure function and the $g$ function, we apply a prior of $X>0$. We check the posteriors and find that this has a marginal effect on the final result.}, including the BAO distance measurements\footnote{Note that galaxy surveys provide BAO distance measurements of $D_{\rm A}/r_{\dd}, H r_{\dd}$ or $D_{\rm V}/r_{\dd}$, instead of $D_{\rm A}, H$ or $D_{\rm V}$, but this does not matter since the unknown amplitude $r_{\dd}$ can be absorbed into $\alpha_0$ or $\beta_0$. As we only use the shape information of $f(a)$ for dark energy studies, the value of $\alpha_0$ or $\beta_0$ is actually irrelevant.}, the luminosity distance measurements from SN Ia observations, and $H(z)$ measurements from the cosmic chronometers \citep{OHD}. 

\section{Simulated datasets and parameter estimation}

\begin{figure}[ht!]
\centering
\includegraphics[width=0.6\linewidth]{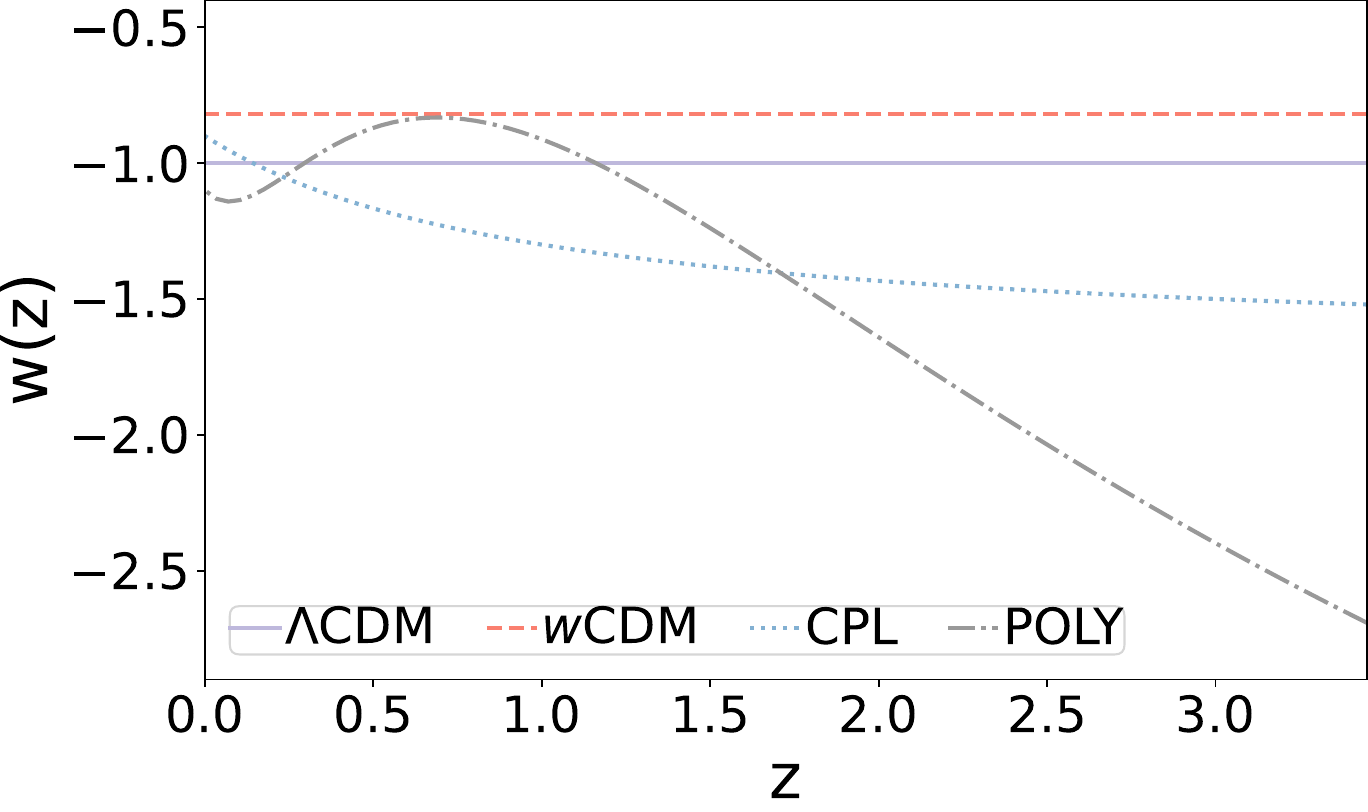}
\caption{The $w(z)$ models used as fiducial models in this work.}
\label{fig:w_model}
\end{figure}

\begin{table}
\begin{center}
\begin{tabular}{c|c}
\hline
\hline
  {Dark energy models} & $w(a)$   \\
\hline
 {$\Lambda$CDM} & $w = -1$          \\ 
 {$w$CDM} & $w = -0.82$          \\ 
 { CPL }         & $w = -0.9 - 0.8(1-a)$                \\
 { POLY }         & $w = -1.1 -1.3(1-a) + 11.2(1-a)^2 -15.7(1-a)^3 $                 \\
\hline
\hline
\end{tabular}
\end{center}
\caption{Dark energy models used for producing the mock datasets.}
\label{table:w}
\end{table}

In this section, we present the simulated datasets used for this work, including the mock BAO datasets for the complete DESI survey, and the mock SNe Ia datasets assuming a sensitivity of the LSST and Roman surveys.

To start with, we choose four phenomenological dark energy models as fiducial models shown in Table \ref{table:w} and Fig. \ref{fig:w_model}. These include the $\Lambda$CDM model ($w=-1$), a $w$CDM model with $w=-0.82$, a Chevallier-Polarski-Linder (CPL) model \citep{Chevallier:2000qy,Linder:2002et} with $w_0=-0.9$ and $w_a=-0.8$, and a more complicated model of $w(a)$, which is a polynomial of $(1-a)$ (POLY). To be generic, the parameters for the CPL and POLY models are chosen so that $w(a)$ is allowed to cross the $w=-1$ boundary, as motivated by observations \citep{quintom, DErecon6, DErecon7, DErecon8, DESI:2024mwx}. Throughout the paper we assume fiducial values of $\Omega_{\M}=0.315$ and $H_0=67.4~{\rm km}~{\rm s}^{-1}~{\rm Mpc}^{-1}$, which are consistent with values in a $\Lambda$CDM model favored by Planck observations \citep{Planck18}. Given the input dark energy models and fiducial values of $\Omega_{\M}$ and $H_0$, the simulated BAO and SNe Ia observables, including the mean values and data covariance matrix, can be created, if the sensitivity of the relevant surveys is assumed. 

\subsection{Simulated BAO observables}
For the BAO observables, we assume a sensitivity of the complete DESI survey. DESI is a Stage IV ground-based galaxy spectroscopic survey, measuring the expansion rate and the growth rate of cosmic structures at (sub)percent level across a wide range of redshifts. We follow the official DESI specifications \citep{2023arXiv230606307D}, and use the forecast sensitivity of $D_{\rm A}/r_{\dd}$ and $H r_{\dd}$ derived from tracers including the Bright Galaxy Samples (BGS), Luminous Red Galaxies (LRGs), Emission Line Galaxies (ELGs), Quasars (QSOs) and the Lyman-$\alpha$ forest (Ly$\alpha$) at $0<z<3.5$.

\subsection{Simulated SNe Ia observables}

Type Ia supernovae, as cosmic standard candles, offer measurements of luminosity distances at multiple redshifts. LSST \citep{LSST} and Roman \citep{Roman} are two main forthcoming SNe Ia surveys with complementary redshift coverage, namely, LSST aims to observe hundreds of thousands SNe at low and intermediate redshifts, while Roman is expected to detect SNe up to $z=3$. We assume that the uncertainties of the SNe distance modulus are quadratic sums of the intrinsic scatter of $\sigma_{\rm int}=0.13$ mag and both the lensing-induced scatter and the peculiar velocity scatter\footnote{We assume ideal cases with no systematics in this simulation.}. For the expected number of SNe to be detected by LSST, we follow \citet{2023arXiv231117176M} to assume a $10$-year survey over $18,000$ ${\rm deg}^2$ with a $15\%$ completeness at $z<0.7$, and for Roman, we assume a WIDE survey mode \citep{Rose:2021nzt}. 

\subsection{Parameter estimation}

Given the simulated datasets, which include a data vector storing the mean value of the observables, and a data covariance matrix, we perform a Monte Carlo Markov Chain (MCMC) analysis to constrain the $\alpha$ parameters defined in Eq. (\ref{Eq:para-DA}) using the {\tt Cobaya} software \citep{Torrado_2021}.

\section{Results}

\begin{figure}[ht!]
\centering
\includegraphics[width=\linewidth]{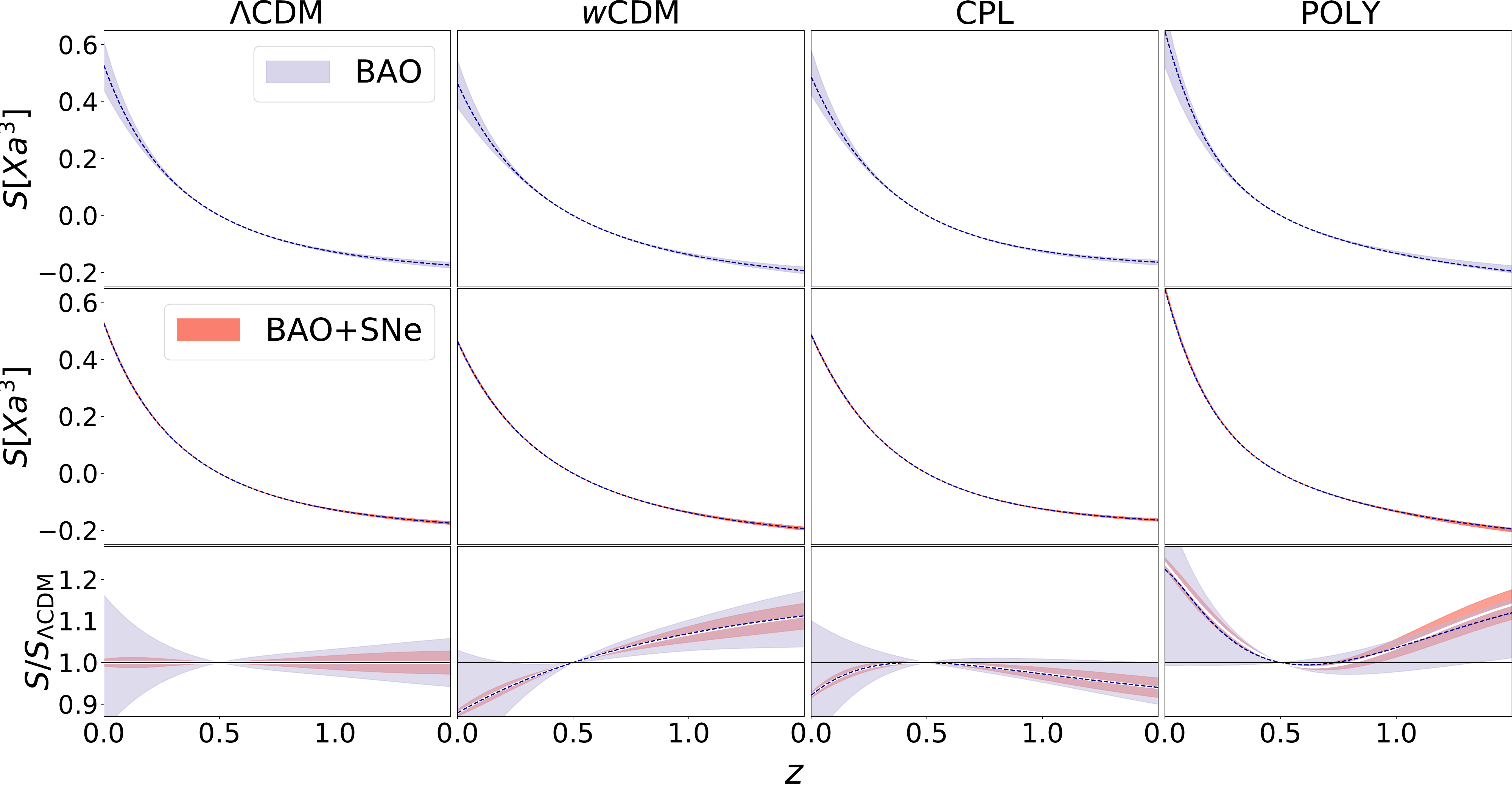}
\caption{
The reconstructed shape function of dark energy, $S[Xa^{3}]$, as a function of redshift $z$ for four input $w(z)$ models as illustrated in the legend. In all panels, the blue dashed lines denote the input model, and the shaded bands show the 68\% confidence level (CL) reconstruction. The top and middle panels present results derived from the simulated BAO assuming a complete DESI survey (blue bands) and BAO combined with SNe datasets assuming complete LSST and Roman surveys (red bands), respectively. The bottom panel shows the result normalised by the $\Lambda$CDM model $S_{\rm \Lambda CDM}$. The white curves in the middle show the mean of the reconstructed $S$ function.}
\label{fig:nf_rec}
\end{figure}

\begin{figure}[ht!]
\centering
\includegraphics[width=1\linewidth]{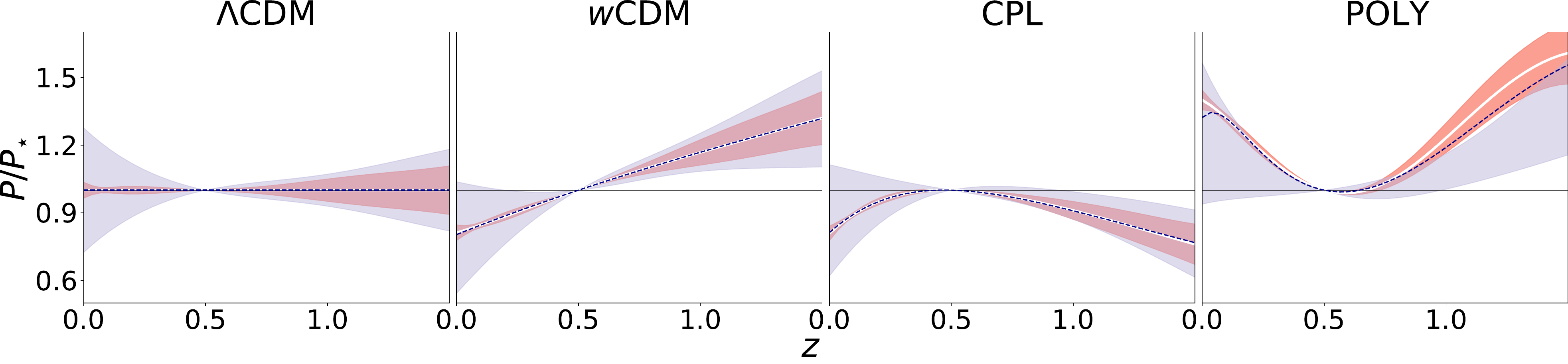}
\caption{The reconstructed pressure function of dark energy, normalised at $a=a_{\star}$, as a function of redshift $z$ for four input $w(z)$ models as illustrated in the legend. In all panels, the blue dashed lines denote the input model, and the shaded bands show the 68\% confidence level (CL) reconstruction derived from the simulated BAO (blue bands) and BAO combined with SNe datasets (red bands), respectively. The horizontal black lines show the $\Lambda$CDM prediction of $P/P_{\star}=1$ for a reference.}
\label{fig:p_rec}
\end{figure}

\begin{figure}[ht!]
\centering
\includegraphics[width=1\linewidth]{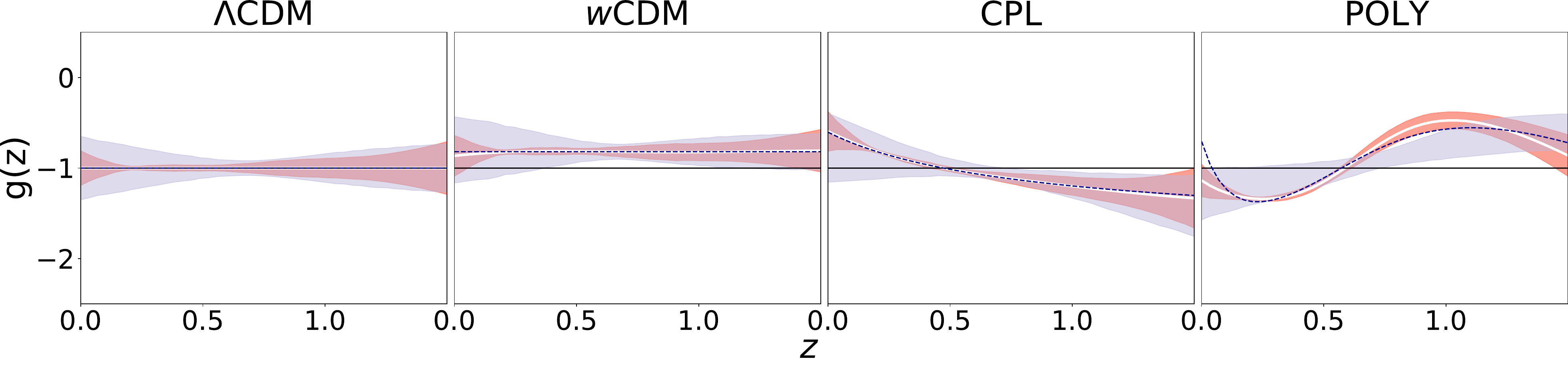}
\caption{Same as Fig. \ref{fig:p_rec}, but for $g(z)$. The horizontal black lines show the $\Lambda$CDM prediction of $g=-1$ for a reference.}
\label{fig:g_rec}
\end{figure}

\begin{figure}[ht!]
\centering
\includegraphics[width=1\linewidth]{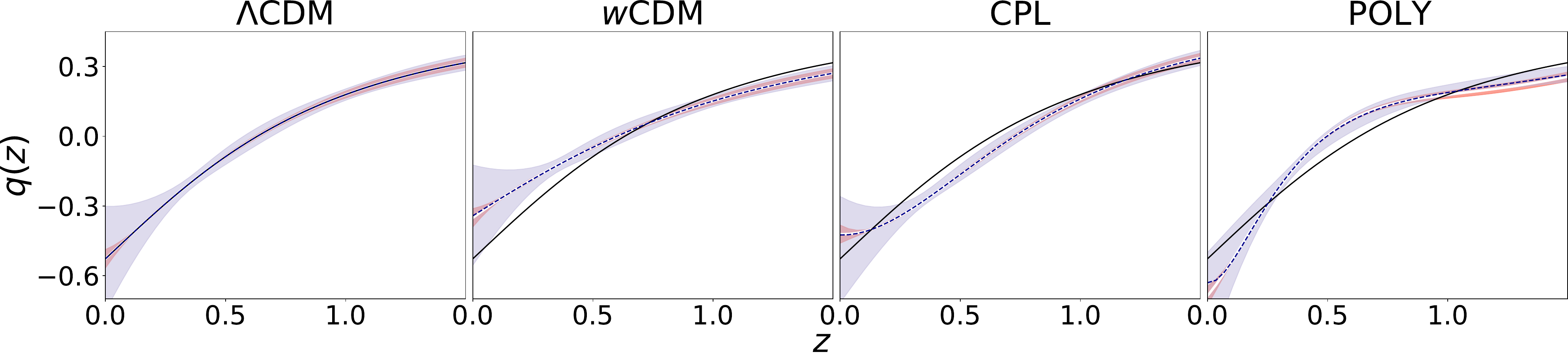}
\caption{Same as Fig. \ref{fig:g_rec}, but for the decelaration function $q(z)$. The black lines show the $\Lambda$CDM prediction for a reference.}
\label{fig:q_rec}
\end{figure}

In this section, we present the main result of this work, as summerised in Figs. \ref{fig:nf_rec}-\ref{fig:q_rec}. 

Fig. \ref{fig:nf_rec} shows the reconstructed shape function of $Xa^3$, derived from the simulated DESI BAO data alone (top panels) and from the simulated DESI BAO + LSST (SNe) + Roman (SNe) (middle panels). For a better visualisation, we show the result normalised by $S$ predicted by the $\Lambda$CDM model in the bottom panel. As shown, the input models (shown in dashed blue curves) are well recovered in all cases, validating our pipeline for reconstructing $S[Xa^3]$ from data. We find that forthcoming BAO combined with SNe observations can well constrain the shape function of dark energy, making it possible to use this function to test the $\Lambda$CDM model to a high precision. 

Figs. \ref{fig:p_rec} - \ref{fig:q_rec} show the reconstructed pressure function $P(z)$ of dark energy (normalised by the value of $P$ at $a_{\star}$), the characterisation function $g(z)$, and the deceleration function $q(z)$, respectively. As shown, all input models can be recovered at $z\gtrsim0.1$ within the small uncertainties derived from the simulated datasets of BAO (DESI) + SNe (LSST+Roman). The POLY model is less well reproduced at $z\lesssim0.1$ due to its wiggly feature at such low redshifts, which requires higher-order terms in the expansion in Eq. (\ref{Eq:para-DA}). We have tried an expansion of Eq. (\ref{Eq:para-DA}) with the $\alpha_7$ term (see Sec. \ref{sec:order}), which indeed improves the accuracy of the reconstruction, but with much larger uncertainties. Therefore our default choice (keeping up to the $\alpha_6$ term) is a reasonable compromise.

\section{Conclusion and Discussions}

In this era of precision cosmology, we have been gaining access to high quality observational data probing the Universe from various angles, which is deepening our understanding of the cosmos. Revealing the nature of dark energy is one of the most challenging problems to tackle in modern cosmology. As dark energy is the driving force of the current acceleration of the spacetime expansion, it is crucial to develop methods and tools to capture critical features of dark energy from measurements of the cosmic expansion.

In this work, we develop methods to extract important features of dark energy, including the shape function of the energy of dark energy $S[Xa^3]$, the evolution history of the pressure $P(a)/P(a_{\star})$, the characterisation function $g(a)$, and the deceleration function $q(a)$, from cosmic distance measurements, primarily including BAO and SNe measurements. We apply our pipeline to simulated DESI, LSST and Roman datasets created for a range of phenomenological dark energy models, and find that our method can well capture the dynamical features of dark energy hidden in the simulated datasets. As our method only extracts information of dark energy from distance measurements, it is by design free from degeneracies among other cosmological parameters. This allows for diagnostic dark energy tests using individual type of observational data, which is important for dark energy studies.

Our method is directly applicable to existing cosmic distance measurements for dark energy tests, which is released in a companion paper \citep{XMW24}. 

\section{acknowledgments}
We thank Ruiyang Zhao for helpful discussions. All authors are supported by the National Key R \& D Program of China (2023YFA1607800, 2023YFA1607803), NSFC grants (No. 11925303 and 11890691), and by a CAS Project for Young Scientists in Basic Research (No. YSBR-092). SY is also supported by a NSFC grant (No. 12203062). SY and GBZ are also supported by science research grants from the China Manned Space Project with No. CMS-CSST-2021-B01. GBZ is also supported by the New Cornerstone Science Foundation through the XPLORER prize.

\newpage

\appendix

\section{DERIVATION OF THE CHARACTERISATION FUNCTION}
\label{sec:g}

By definition,
\begin{equation}
    f(a) \equiv H^{2}a^{3} = H^{2}_{0}\left[\Omega_{\M} +(1-\Omega_{\M})X(a)a^{3} \right],
\end{equation}
Then
\begin{equation}\label{f-der}
    \begin{aligned}
        f^{\prime}(a) & = \frac{\dd}{{\dd} a}H^{2}_{0}\left[\Omega_{\M} +(1-\Omega_{\M})X(a)a^{3} \right] 
        = -3H_{0}^{2}(1-\Omega_{\M})w(a)X(a)a^{2},
    \end{aligned}
\end{equation}
\begin{equation}
    \begin{aligned}
        f^{\prime\prime}(a) &= -3H_{0}^{2}(1-\Omega_{\M})\frac{\dd}{{\dd}a}\left[w(a)X(a)a^{2}\right] \\
        &= -3H_{0}^{2}(1-\Omega_{\M})\left[ w^{\prime}(a)X(a)a^{2} + w(a)X(a)(-3)\frac{1+w(a)}{a}a^{2} + w(a)X(a)2a \right]\\
        &= -3H_{0}^{2}(1-\Omega_{\M})\left[a^{2}w^{\prime}(a)X(a) - aw(a)X(a) - 3aw^{2}(a)X(a)\right],
    \end{aligned}
\end{equation}
Thus
\begin{equation}
    \begin{aligned}
        \frac{f^{\prime\prime}(a)}{f^{\prime}(a)} = \frac{1}{a}\left[a\frac{w^{\prime}(a)}{w(a)}-3w(a) - 1\right].
    \end{aligned}
\end{equation}
Hence,
\begin{equation}
    g(a) \equiv   w-\frac{a}{3} \frac{w^{\prime}}{w} =  -\frac{1}{3}\left[a\frac{f^{\prime\prime}(a)}{f^{\prime}(a)} + 1\right].
\end{equation}

\section{Choice of the pivot point and order of the expansion} 
\label{sec:zp_order}

\subsection{Impact of the choice of $z_{\rm p}$ on the reconstructed $f$}
\label{sec:zp}

\begin{figure}[ht!]
\centering
\includegraphics[width=1\linewidth]{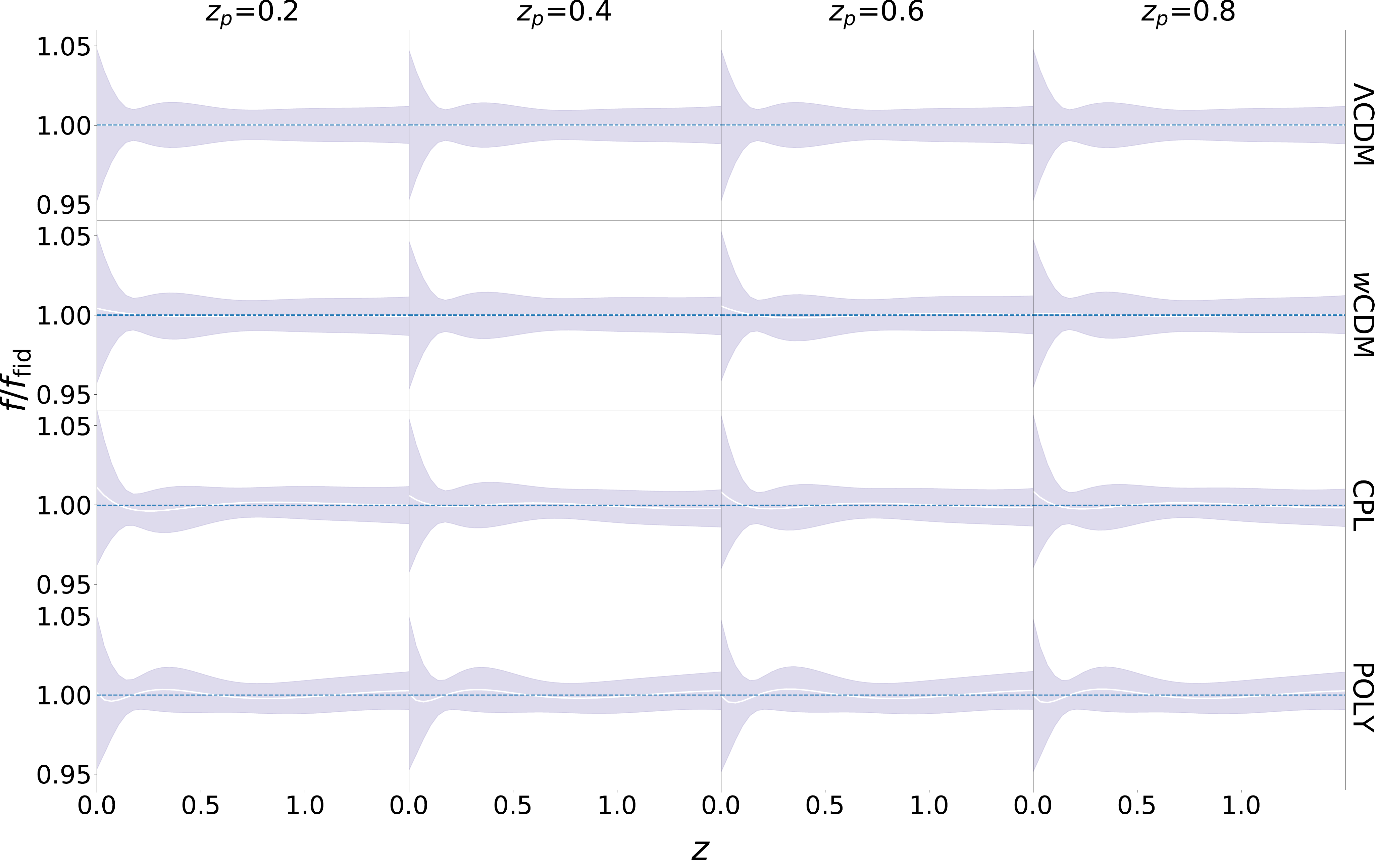}
\caption{The reconstructed dark energy shape function $f$ normalised by the input fiducial function for four different choices of $z_{\rm p}$. The mean (white lines) and 68\% CL uncertainties (shaded regions) derived from the simulated DESI data are shown for four phenomenological dark energy models illustrated in the legend. The horizon blue dashed lines show $f/f_{\fid}=1$ for a reference.
}
\label{fig:test_zp}
\end{figure}

We test how the choice of $z_{\rm p}$ affects the reconstructed dark energy shape function $f$, as defined in Eq. (\ref{eq:fNorm}). For this test, we reconstruct $f$ from the simulated DESI data created 
with various dark energy models with four choices of $z_{\rm p}$ of $0.2,0.4,0.6$ and $0.8$, and show the result in Fig. \ref{fig:test_zp}. As expected, the choice of $z_{\rm p}$ has negligible effect on the final reconstructed $f$, which demonstrates the robustness of our reconstruction result.

\subsection{Impact of the expansion order}
\label{sec:order}

\begin{figure}[ht!]
\centering
\includegraphics[width=0.8\linewidth]{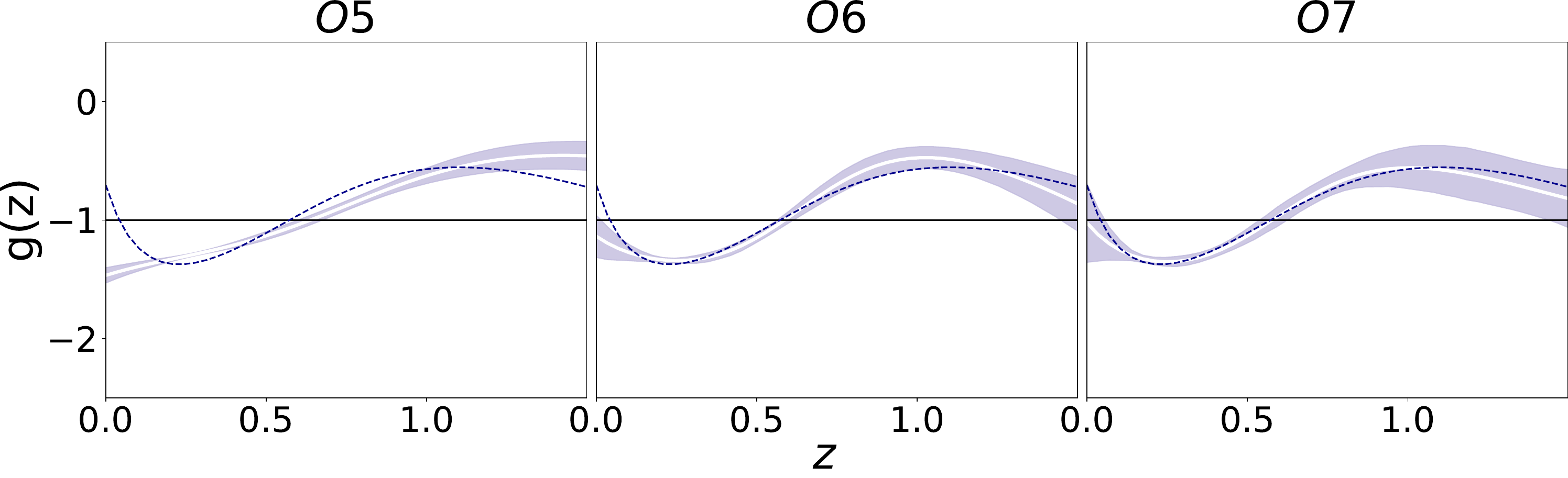}
\caption{The mean (white lines) and 68\% CL uncertainties (shaded regions) of the reconstructed $g$ function of the POLY fiducial model using expansions up to the $x^5$ ($O5$), $x^6$ ($O6$) and $x^7$ ($O7$) terms in Eq. (\ref{eq:chi_order}). The blue dashed lines show the input fiducial model, and the black horizon lines show the $\Lambda$CDM prediction for a reference.}
\label{fig:gO567}
\end{figure}

The maximum order of the expansion in Eq. (\ref{Eq:para-DA}) does affect the final reconstruction of the functions that are closely related to dark energy. As a demonstration, in Fig. \ref{fig:gO567} we show the reconstructed $g(a)$ from the simulated DESI+LSST+Roman datasets for the POLY model with different maximum orders of expansion, namely, we use expansions up to the $x^5$ ($O5$), $x^6$ ($O6$) and $x^7$ ($O7$) terms in Eq. (\ref{eq:chi_order}), respectively. As shown, the reconstruction bias is significant in the $O5$ case, while the uncertainties are large in the $O7$ case, making the $O6$ case a reasonable compromise, which is used for producing main results of this work.

\section{DERIVATION OF THE
DISTANCE-REDSHIFT PARAMETERS}
\label{sec:expansion}

To be general, we expand $\chi/\chi_{\rm fid}$ up to order $N$, namely,
\begin{equation}\label{eq:chi_order}
\chi(z)=\alpha_0\left( 1 + \sum_{i=1}^{N}\frac{1}{i!}\alpha_{i}x^{i}\right) \chi_{\rm fid}(z)
\end{equation}
where
\begin{equation}
x \equiv \chi_{\fid}(z) / \chi_{\fid}(z_{\mathrm{p}})-1
\end{equation}
Taking derivative of Eq. (\ref{eq:chi_order}) with respect to $z$, we have
\begin{equation}
\begin{aligned}
    \frac{1}{H(z)} & = \frac{\dd}{{\dd}z} \chi(z) 
    = \alpha_0\left[
    \chi_{\fid}(z)\left( \sum_{i=1}^{N}\frac{1}{(i-1)!}\alpha_{i}x^{i-1} \right) \frac{{\dd}x}{{\dd}z} +
    \left(1+\sum_{i=1}^{N}\frac{1}{i!}\alpha_{i}x^{i}\right)\frac{\dd}{{\dd}z}\chi_{\fid}(z)\right]\\
    & = \alpha_0\left[
    \chi_{\fid}(z)\left(\sum_{i=1}^{N}\frac{1}{(i-1)!}\alpha_{i}x^{i-1} \right) \frac{1}{H_{\fid}(z) \chi_{\fid}(z_{\mathrm{p}})}+
    \left(1+ \sum_{i=1}^{N}\frac{1}{i!}\alpha_{i}x^{i}\right)\frac{1}{H_{\fid}(z)}\right]\\
\end{aligned}
\end{equation}
Then
\begin{equation}
    \begin{aligned}
        \frac{H_{\fid}(z)}{H(z)} &= \alpha_0\left[
        \left(\sum_{i=1}^{N}\frac{1}{(i-1)!}\alpha_{i}x^{i-1}\right) \frac{\chi_{\fid}(z)}{\chi_{\fid}(z_{\rm p})}+
        \left(1+ \sum_{i=1}^{N}\frac{1}{i!}\alpha_{i}x^{i}\right)\right]\\
        &=
        \alpha_0\left[
        \left( \sum_{i=1}^{N}\frac{1}{(i-1)!}\alpha_{i}x^{i-1}\right)(1+x)+
        \left(1+ \sum_{i=1}^{N}\frac{1}{i!}\alpha_{i}x^{i}\right)\right]\\
    \end{aligned}
\end{equation}

For the case of $N = 6$ for example,
\begin{equation}
\frac{D_{\rm A}}{D_{\rm A, fid}} =
\alpha_0\left(1+\alpha_1 x+\frac{1}{2} \alpha_2 x^2+\frac{1}{6} \alpha_3 x^3 +\frac{1}{24}\alpha_4 x^4 + \frac{1}{120}\alpha_5 x^5 +\frac{1}{720}\alpha_6 x^{6} \right) 
\end{equation}

\begin{equation}
    \begin{aligned}
        \frac{H_{\fid}(z)}{H(z)}
        &=
        \alpha_0
        \left(\alpha_1 + \alpha_2 x + \frac{1}{2}\alpha_{3}x^{2} + \frac{1}{6}\alpha_{4}x^{3} + \frac{1}{24}\alpha_5 x^{4} +\frac{1}{120}\alpha_6 x^{5} \right)(1+x) + \\
        &
        \alpha_0\left(1+\alpha_1 x+\frac{1}{2} \alpha_2 x^2 + \frac{1}{6}\alpha_3 x^3 + \frac{1}{24}\alpha_{4}x^{4} + \frac{1}{120}\alpha_5 x^5 +\frac{1}{720}\alpha_6 x^{6}\right)\\
    \end{aligned}
\end{equation}

\newpage
  
\bibliographystyle{raa}
\bibliography{bibtex}

\end{document}